\documentclass[pr,twocolumn]{revtex4}
\usepackage{graphics,graphics,bm,amsmath,mathrsfs,amsfonts,amssymb,latexsym}
\begin{document}

\title{Review of Entangled Coherent States}

\author{Barry C.\ Sanders \\
	\it Institute for Quantum Information Science, University of Calgary, 
	Alberta T2N 1N4, Canada}

\begin{abstract}
We review entangled coherent state research since its first implicit use in 1967 to the present.
Entangled coherent states are important to quantum superselection principles, quantum information processing, quantum optics, and mathematical physics.
Despite their inherent fragility they have produced in a conditional propagating-wave quantum optics realization.
Fundamentally the states are intriguing because they are entanglements of the coherent states, which are in a sense the most classical of all states of a dynamical system.
\end{abstract}
\pacs{03.67.Bg,42.50.Dv,42.50.St}
\maketitle
\tableofcontents
\section{Introduction}
\label{sec:introduction}

Coherent states play an important role in representing quantum dynamics, particularly when
the quantum evolution is close to classical. The coherent state was originally introduced by
Schr\"{o}dinger in 1926 as a Gaussian wavepacket to describe the evolution of a harmonic
oscillator~\cite{Sch26}. This centroid of the distribution obtained from the Gaussian wavefunction followed the classical evolution of the harmonic oscillator, thereby mimicking its period evolution,
and the spread of the wavepacket was fixed. Furthermore the spread of this wavefunction 
satisfied the Heisenberg uncertainty relation and hence was as localized as possible within
the requirements of quantum theory.

The coherent state emerged as an important representation with the advent of the laser and
the concomitant desire to juxtapose quantum electrodynamics with analyses of coherent optical
systems. As the electromagnetic field in free space can be regarded as a superposition of
many classical modes, each one governed by the equations of a simple harmonic oscillator,
the coherent state became significant as the tool for connecting quantum and classical optics.

The coherent state in quantum optics embodies the quantum-to-classical transition.
Coherent states are minimum-uncertainty states.
The centroid (mean values of the canonical variables) follows the evolution of the classical canonical variables in the classical optical
description.
In addition coherent states are eigenstates of the annihilation operator, hence correspond to 
classical noiseless fields in direct detection by ideal point electric-dipole detectors.
Thus coherent states play an important role in quantum optics by connecting classical
and quantum descriptions of quantum optical phenomena.

As coherent states are regarded as quasiclassical, the introduction of superpositions of
coherent states was rapidly of widespread interest. Evidence of such superpositions first
appeared in a study of a certain type of nonlinear Hamiltonian evolution by Milburn~\cite{Mil85,MH86},
and the manifestation of superpositions of coherent states was analyzed in detail
by Yurke and Stoler~\cite{YS86,YS87}. Studies of superpositions of coherent states for a single
mode of the electromagnetic field concerned how to produce such states, their properties
(such as squeezing, photon number distribution, and robustness to environmental decoherence),
and extensions to generalized coherent states~\cite{TG65,Bia68,Sto71,BG71,ACGT72,Gil72,Per72,Per86}.

Superpositions of coherent states have been reviewed by Bu\v{z}ek and
Knight~\cite{BK95}. Superpositions of nearly distinct coherent states earned the term
`cat state', in deference to Schr\"{o}dinger's paradox of the cat, whose state of existence
seems to be in a superposition of being dead vs alive~\cite{Sch35}. As the states of death and life are
considered to be macroscopically valid and distinct, the superposition of two coherent states,
with large amplitude phases separated by $\pi$ radians, is analogous to this paradox.

Superpositions of coherent states are difficult to produce, and fundamentally this could
be due to extreme sensitivity to environmental decoherence. In fact this sensitivity 
is important in informing us as to why such peculiar states are prevalent in nature. Experimental
efforts to create cat states have concentrated on creating superpositions of coherent states
that have limited distinguishability~\cite{BHD+96}. 
Such states have been dubbed Schr\"{o}dinger kittens~\cite{OTLG06}.

Soon after the introduction of these single-mode superpositions of coherent states, 
entangled coherent states (or superpositions of multimode coherent states) became of
widespread interest. These superpositions of multimode coherent states arose 
independently in several papers. The earliest entangled coherent state appears in Eq.~(10) of the 1967
Aharonov and Susskind charge-superselection analysis that shows charges could appear in superposition~\cite{AS67}.
Entangled coherent states provided a representation by which superpositions of charge states could be understood 
within the confines of superselection~\cite{SBRK03}.

The next appearance of the entangled coherent state appeared in Eq.~(11) of Yurke and Stoler's 1986 seminal paper on 
generating superpositions of coherent states~\cite{YS86}. They showed that 
mixing a superposition of coherent states with a vacuum state at a beam splitter
yields an entangled coherent state output.  Subjecting one of the two output field modes to homodyne
detection~\cite{YS80,YC83,TS04}, which corresponds to a quadrature measurement~\cite{YS86,YS87,SLK95}.

The entangled coherent state representation appeared again in considerations of
pair coherent states by Agarwal~\cite{Aga86,Aga88,AT91}:
the entangled coherent state representation of the pair coherent state appears in Eq.~(2.6) of Agarwal's 1988 work~\cite{Aga88}
Although pair coherent states,
which are a special case of the Barut and Girardello SO(2,1)$\simeq$SU(1,1)$\simeq$SL(2,$\mathbb R$) coherent state~\cite{BG71},
are the objects of study, these pair coherent states are elegantly represented as
continuous bipartite entangled coherent states.

Entangled coherent states as entities of physical interest in their own right,
first arose in a study by Tombesi and Mecozzi~\cite{MT87,TM87}, where they generalize
the nonlinear birefringent evolution of Milburn~\cite{Mil85, MH86} and Yurke and Stoler~\cite{YS86}
to multimode coherent states.
Rather than employ the single-mode nonlinear evolution
associated with the ideal optical Kerr nonlinearity, they treat the ideal Hamiltonian
evolution of two orthogonally polarized light beams interacting in a nonlinear
birefringent medium. After an appropriate evolution time, an initial 
two-mode coherent state evolves to an entangled coherent state~\cite{MT87,TM87,SR99,SR00}.

Tombesi and Mecozzi then studied the 
relevant statistics of these states, such as photon number distribution and squeezing,
as well as robustness to decoherence~\cite{TM87}. Agarwal and Puri~\cite{AP89}
studied evolution of a two-mode coherent state through an optical Kerr medium
and studied the entangled coherent state and its statistical properties. They also pointed
out that their entangled coherent state is a simultaneous eigenstate of operators
that are quadratic in the annihilation operators for the two modes.

The term `entangled coherent state' was introduced by Sanders in a study concerning production
of entangled coherent states by using a nonlinear Mach-Zehnder interferometer~\cite{San92,San92E}.
The nonlinear interferometer comprises a nonlinear medium in one path of a Mach-Zehnder interferometer.
The nonlinear medium alone could suffice to produce entangled coherent states~\cite{MT87,TM87,SR99,SR00},
but the interferometric set-up has analogies with the homodyne detection concept for superpositions of coherent states~\cite{YS86}.
Linear optics alone is known to be insufficient to generate entangled coherent states so a nonlinearity is required~\cite{HB08}.

Soon after the proposals to create entangled coherent states in two-mode propagating fields, 
a cavity quantum electrodynamics realization was proposed using one atom traversing two cavities and post-selecting on atomic measurement.
This scheme was suggested for realizing entanglement between a coherent state in one mode and the vacuum in the other mode~\cite{DMM+93}.

Entanglement of a coherent state with a vacuum
state (which is a coherent state of zero amplitude) was a particular focus of the Sanders proposal~\cite{San92}.
In this analysis, a bipartite entangled coherent states was shown to violate a 
phase-coherence Bell inequality~\cite{GPY88} in the few-photon limit~\cite{San92,San92E}.
Later the entangled coherent states were also shown to violate a formal 
Bell inequality in the large photon number limit~\cite{MSM95}.

Coherent states generated by a Kerr nonlinearity, within or without an interferometer,
obey a conservation rule for total photon number, which constrains the phase relationship
between the components of the multimode coherent state superposition. Chai~\cite{Cha92}
introduced entangled coherent states as superpositions of two-mode coherent states
with equal amplitude but opposite in optical phase and allowed an arbitrary phase
relationship between the two components of the bipartite superposition. The analysis then
focused on the two-mode extension to single-mode even and odd coherent
states~\cite{DMM74}, which are sometimes called `even entangled coherent states' and
`odd entangled coherent states'~\cite{DMN95}.
These even and odd entangled coherent states naturally generally for $q$-coherent states~\cite{Spi95,Spi96E}.

Chai studied statistical properties of even
and odd entangled coherent states and
showed that the joint photon number distribution of such entangled coherent states
vanished for certain values of the photon number sum or difference. He also
evaluated the squeezing properties of these even
and odd entangled coherent states.
These even and odd entangled coherent states have quantum metrological applications~\cite{ADM+94,AM94}.
Such states could be constructed by multimode parametric amplifiers~\cite{DMN95}.
Entangled coherent states also have applications to quantum information processing~\cite{MMS00}

Although `balanced', or equally weighted, superpositions of multimode coherent states are typically studied,
`unbalanced', or unequally weighted, superpositions are possible.
An approximation to ideal unequally weighted superpositions
can be generated by a nonlinear evolution within a double cavity system~\cite{WS93}.
The requisite nonlinear evolution is actually a special case of the nonlinear evolution that leads to
Titulaer-Glauber generalized coherent states~\cite{TG65,Bia68,Sto71}.

Entangled coherent states were initially treated as bimodal states but later generalized to 
superpositions of multimode coherent states~\cite{JTS95,Zhe98,WS01}.
Generalizations to multimode systems allows the intricacies of multipsartite entanglement to become manifest in 
entangled coherent states, for example in Greenberger-Horne-Zeilinger and W types of states~\cite{JA06,LY09}
and entangled coherent state versions of cluster states~\cite{MSVR08,BCAB08, WSL08}.

As entangled coherent states exhibit entanglement, which is a resource for quantum
information protocols, entangled coherent states have been studied both as a resource
for quantum information protocols and also as an input to a quantum information protocol.
The degree of entanglement embodied by entangled coherent states was studied in the
context of quantum information, where entanglement is considered a resource.

By showing that the even bipartite entangled coherent state can be obtained by mixing an
even coherent state with the vacuum at a beam splitter, van Enk and Hirota~\cite{vEH01}
establish that the even entangled coherent state has precisely one ebit of entanglement,
where one ebit is the amount of entanglement in a maximally entangled state of
two qubits, or spin-$\tfrac{1}{2}$ particles. The degree of entanglement in a bipartite
entangled coherent state generated by a nonlinear Kerr evolution was subsequently
shown to yield an arbitrarily large amount of entanglement over 
proportionately short times and a limited amount of entanglement over longer times~\cite{vEn03}.

Entangled coherent states have been employed in quantum teleportation tasks~\cite{BBC+93} in two ways:
as the state being teleported via continuous-variable quantum teleportation~\cite{Wan01,JBS02}
and as the entangled resource state employed to effect the teleportation~\cite{Wan01,vEH01,JKL01,JBK+02}.
Teleportation gives operational meaning to the amount of entanglement in an entangled coherent state
as teleportation consumes prior shared entanglement to transport quantum information through a classical channel~\cite{BBC+93}.

Entangled coherent states can go beyond entangling harmonic oscillator coherent states.
Earlier in this section we mentioned the Barut-Girardello coherent states~\cite{BG71},
which can be used to construct pair coherent states,
and these states are also an example of generalizing coherent states beyond the Heisenberg-Weyl algebra.
Gilmore and Perelomov independently showed another way of generalizing coherent states based on abstracting the displacement operator to general group operations
acting on minimum- or maximum-weight states~\cite{ACGT72,Gil72,Per72,Per86}.
The orbit of Gilmore-Perelomov states under the general group action form the coherent states for the given group.
For example entangled coherent states can be constructed as superpositions of tensor products
of two or more generalized Perelomov or Barut-Girardello $\mathfrak{su}(2)$~\cite{WSP00,GBHA08} and $\mathfrak{su}(1,1)$ coherent states as well as entangled binomial states~\cite{WSP00}.

In fact entangled coherent states arise naturally from the nonclassical coalgebraic structure of the generalized boson algebra ${{\cal U}}_{\langle q\rangle}(h(1))$~\cite{ACS05}.
Squeezed states~\cite{LK87} are a generalization of coherent states as orbits of squeezed vacuum states under the Heisenberg-Weyl displacement operator.
Specifically squeezed states are constructed as orbits of the squeezed vacuum state
whereas coherent states are obtained by the same orbit construction but with the vacuum replacing the squeezed vacuum state as the fiducial state~\cite{SLK95}.
Another way of dealing with squeezing in the context of entangled coherent states to is to subject them to squeezing~\cite{LJ09} or to use squeezed
light in the homodyne detection process to enhance detector performance~\cite{TM87}.

Generalized beyond the coherent state framework is also also useful.
For example coherent states can have photons added to them thereby creating `photon-added coherent states'~\cite{AT91},
which leads naturally to entangled photon-added entangled coherent states~\cite{Tao02,ZX09}.
The ``single-mode excited entangled coherent states'' also involve adding a photon by acting on an entangled coherent state directly with a photon-creation operator~\cite{XK06,RJZ08},
and this state has value as a cyclic representation of the $\mathfrak{hw}$ algebra~\cite{XK06}.

This review provides an overview of research into entangled coherent states and their generalizations, implementations and applications.
This field of research is quite large so not every paper is cited, but this review strives to be comprehensive in covering all the directions 
concerning entangled coherent states.
With the recent successful generation of entangled coherent states~\cite{OFTG09} and their potential importance in quantum information processing,
many new discoveries can be expected in the near future.

\section{Formalism}
\label{sec:formalism}

\subsection{Coherent states of the simple harmonic oscillator}
\label{subsec:simple}

Coherent states of the simple harmonic oscillator are well known since the foundational work of Schr\"{o}dinger~\cite{Sch26} and 
the ubiquity of coherent states in quantum optics~\cite{Gla63a,Sud63,Kla85}.
The Hamiltonian for the quantized simple harmonic
oscillator in one dimension is
\begin{equation}
\hat{H}=\frac{\hat{p}^2}{2m}+\frac{1}{2}m\omega^2\hat{q}^2
\end{equation}
with $\hat{}$ signifying an operator, $m$ the mass of the oscillator, $\omega$ the
angular frequency, and $\hat{q}$ and $\hat{p}$ the canonically conjugate Hermitian
operators for position and momentum, respectively. 
These conjugate operators satisfy the commutator relation
$[\hat{q},\hat{p}]=i\hbar$, and the Hamiltonian spectrum is $(n+\frac{1}{2})\hbar\omega$
for $n$ a nonnegative integer. The harmonic oscillator thus has a ground state energy level
of $\frac{1}{2}\hbar\omega$ and all excited energy levels are integer multiples of
$\hbar\omega$ above the ground state energy level. For the simple harmonic oscillator,
$n$ indicates the number of phonons, with each additional phonon increasing the 
oscillator's mechanical energy by $\hbar\omega$.

As the energy levels are equally spaced, it is convenient to introduce
the phonon lowering operator
\begin{equation}
	\hat{a}^\dagger=\sqrt{\frac{m\omega}{2\hbar}}\hat{q}
		+\frac{i\hat{p}}{\sqrt{2m\hbar\omega}}
\end{equation}
and its conjugate raising operator $\hat{a}^\dagger$,
which satisfy the commutator relation
\begin{equation}
\label{eq:aadagger}
	[\hat{a},\hat{a}^\dagger]=\openone
\end{equation}
corresponding to the Heisenberg-Weyl Lie algebra $\mathfrak{hw}(1)$ comprising generators
$\{\hat{a},\hat{a}^\dagger,\openone\}$ with~$\openone$ the identity operator.

The Fock number states~$|n\rangle$, for $n$ the number of phonons,
provide a countable, orthonormal basis for the Hilbert space~$\mathscr{H}$,
with $n$ the number of phonons and $|n\rangle$ an eigenstate of the number
operator:
\begin{equation}
\label{eq:numberop}
	\hat{n}\equiv\hat{a}^\dagger\hat{a},\;\hat{n}|n\rangle=n|n\rangle.
\end{equation}
For the extension to $n$ coupled simple harmonic oscillators, each indexed by
an integer $\ell$, the algebra for the identity~$\openone$ and the $2n$ operators is 
\begin{equation}
\label{eq:hwn}
	\{\hat{a}_\ell,\hat{a}_\ell^\dagger; \ell=1,2,\ldots,n\}, \;
	[\hat{a}_\ell,\hat{a}^\dagger_{\ell'}]=\delta_{\ell\ell'}\openone.
\end{equation}

The number states are energy eigenstates, hence are stationary states of the simple
harmonic oscillator. An alternative is a Gaussian wavefunction (note that a 
Gaussian in position representation is a Gaussian in the momentum representation
so it is sufficient to refer to a Gaussian wavefunction without specifying position
or momentum representation), which satisfies the Schr\"{o}dinger equation and
is not stationary. The centroid (mean position and momentum) follows the simple
harmonic motion expected for the classical simple harmonic oscillator, and
the Gaussian remains a minimum-uncertainty state (with respect to uncertainty
in position and momentum) so the Gaussian wavefunction has desirable properties.

This Gaussian, with initial conditions such that the position and momentum uncertainties
are stationary, is known as a coherent state due to Glauber's association of such
a state with coherence in quantum optics.
Glauber's oscillating wavepacket is an eigenstate of the annihilation operator~$\hat{a}$
for a given harmonic oscillator. This pure-state wavefunction is given in the
Fock representation by
\begin{equation}
\label{eq:coherentstate}
	\left|\alpha\right\rangle
			= D(\alpha)|0\rangle
			=\exp \left( -\frac{|\alpha|^2}{2}\right)\sum_{n=0}^{\infty} \frac{\alpha^n}{\sqrt{n!}} |n\rangle,
\end{equation}
with $\alpha$ a complex dimensionless amplitude
such that the mean position and momentum are given by 
\begin{equation}
	\bar{q}=\sqrt{\frac{2\hbar}{m\omega}}{\rm Re}(\alpha),\;
	\bar{p}=\sqrt{\frac{\hbar\omega}{2m}}{\rm Im}(\alpha),
\end{equation}
respectively.
The quantity $\bar{n}=|\alpha|^2$ is the mean phonon number,
with the phonon number given by
a Poisson distribution
\begin{equation}
\label{eq:Poisson}
	\Pi_{\bar{n}} = \left|\left\langle n\Big |\alpha=\sqrt{\bar{n}}e^{i\varphi}\right\rangle\right|^2
		=e^{-\bar{n}}\bar{n}^n/n!~
\end{equation}
and 
\begin{equation}
\label{eq:displacement}
	 D(\alpha)=\exp(\alpha\hat{a}^\dagger-\alpha^*\hat{a})
\end{equation}
the displacement operator.
The Poisson distribution has the property that 
\begin{equation}
\label{eq:Poissonmeanvariance}
	\bar{n}=(\Delta n)^2~,
\end{equation}
and the Mandel $Q$ parameter is~\cite{Man79}
\begin{equation}
\label{eq:MandelQ}
	Q=\frac{(\Delta n)^2-\bar{n}}{\bar{n}}~,
\end{equation}
which is unity for the Poisson distribution~(\ref{eq:Poisson}), less
than unity for sub-Poissonian distributions and greater than unity for super-Poissonian
distributions.

The coherent state can be used for multiple harmonic oscillators.
For~$n$ simple harmonic oscillators, the joint coherent state is
\begin{equation}
\label{eq:prodcohstate}
	|\bm{\alpha}\rangle = \prod_{i=1}^N|\alpha_i\rangle
		= D(\bm{\alpha})|\bm{0}\rangle
\end{equation}
with $D(\bm{\alpha})$ a product of single-mode displacement operators~(\ref{eq:displacement})
and $|\bm{0}\rangle$ the joint ground state of all~$n$ oscillators.
As discussed above, a product coherent state with respect to a specific set of modes transforms
to another product coherent state with via a linear mode transformation.

Although this subsection has been concerned with the motion of the simple harmonic
oscillator, and the energy quanta that separate the equally spaced energy levels
have been referred to as phonons, the analysis also applies to the electromagnetic field.
The free-space field is reducible to modes, and the dynamics of each mode corresponds
to a simple harmonic oscillator. The energy quanta are photons, not phonons, as discussed
in subsection~\ref{subsec:quoptics}.

\subsection{Coherent states in quantum optics}
\label{subsec:quoptics}

In quantum optics, the electromagnetic field can be decomposed into modes,
and the dynamics of each mode in free space is equivalent to the dynamics of the simple
harmonic oscillator, with $n$ the number of photons (rather than phonons) in the given mode. 
The canonically conjugate operators~$\hat{q}$ and~$\hat{p}$ are referred to as
the in-phase and out-of-phase quadratures, respectively, with the phase reference
being a local oscillator.

As the phase of the local oscillator can be varied continuously,
it is convenient in quantum optics to define the quadrature operator as
\begin{equation}
	\hat{q}_\theta\equiv\hat{q}\cos\theta+\hat{p}\sin\theta~,
\end{equation}
with $\hat{q}=\hat{q}_0$
and $\hat{p}=\hat{q}_{\pi/2}$. Measurements of a quadrature are performed by 
mixing the signal field with the coherent local oscillator field in an optical homodyne detection
apparatus~\cite{YS80,YC83,TS04}, with the local oscillator field
determining the phase~$\theta$.

In developing a theory of coherence for optical fields,
Glauber employed the coherent state~\cite{Gla63a},
but, instead of the variables being position and momentum of a massive particle in a harmonic
potential, the canonically conjugate 
variables are the in-phase and out-of-phase quadratures of each mode
of the field. The two quadrature field operators are
constructed from the raising and lowering operators for the field mode
with three-vector label~$\vec{k}$ and polarization index~$\varepsilon$,
namely $\hat{a}^\dagger_{\vec{k}\varepsilon}$ and 
$\hat{a}_{\vec{k}\varepsilon}$, respectively. Properties of the coherent state have
been discussed in Subsec.~\ref{subsec:simple} and are investigated in detail by
Klauder and Skagerstam~\cite{Kla85}.

Coherent states have served as a valuable tool for studying quantum optics,
primarily because of the convenience of these states as representations. In addition to a coherent
state being an eigenstate of the annihilation operator~$\hat{a}$, another critical 
property is that the product coherent state is the unique state that transforms to 
a product coherent state under the action of a linear mode coupler such as a beam 
splitter~\cite{AFLP66}. 
For the mode coupler transformation given by
\begin{equation}
	\begin{pmatrix} \hat{a}'_1 \\ \hat{a}'_2 \end{pmatrix}
		=\begin{pmatrix} \cos\theta & e^{i\phi}\sin\theta \\
		-e^{-i\phi}\sin\theta & \cos\theta \end{pmatrix}
		\begin{pmatrix} \hat{a}_1 \\ \hat{a}_2 \end{pmatrix}
\end{equation}
a product coherent state then transforms to a product coherent state.
This property is particularly important in that a coherent state for the field remains
a coherent state under any linear mode transformation.

Whether coherent states can be considered as ontologically real has been the subject
of vigorous debate, both in the context of coherence of atomic
Bose-Einstein condensates~\cite{Mol97, Gea98,Mol98}
and with respect to 
quantum teleportation of coherent states~\cite{RS01,vEF02,SBRK03,Wis04,Smo04}.
The problem essentially concerns the establishment of the phase~$\varphi$ 
of the coherent state either through its creation from a source or by a phase-sensitive
detection of the state.
In practice phase-locking mechanisms
exist that ensure that the phase of the coherent field is correlated with a reference field,
and, treating that field as classical, provides classical meaning to the parameter~$\varphi$.
This issue is discussed in more detail in Subsec.~\ref{subsec:representation}, which is concerned
with the entangled coherent state representation.

Given the indisputable value of the coherent state as a representation, there are two
useful ways to represent the density matrix~$\rho$ of the single-mode field in terms
of coherent states. One makes use of the Glauber-Sudarshan $P$-representation~\cite{Gla63b,Sud63}
and the other makes use of the Husimi distribution, or $Q$ function~\cite{Hus40}
(not to be confused with the Mandel~$Q$~\cite{Man79})
whose advantages with respect to superpositions of coherent states were elaborated
by Milburn~\cite{Mil85}.

The density matrix can be expressed in these representations as
\begin{equation}
\label{eq:PQ}
	\rho=\int \frac{d^2\alpha}{\pi} P(\alpha),\,
	Q(\alpha)=\frac{\langle \alpha|\rho\left|\alpha\right\rangle}{\pi}.
\end{equation}
In quantum optics, the field is said to be semiclassical if $P(\alpha)$ satisfies the axiomatic
requirements of a probability density and is `quantum' otherwise.
In contrast the $Q$-function is always a probability density. There is a continuum of these so-called
quasiprobabilities introduced by Cahill and Glauber~\cite{Cah69},
which are obtained through Gaussian convolution of $P(\alpha)$
with a Gaussian, with $s$ corresponding to $+1$ for $P$, $0$ for the Wigner
function~$W(q,p)$, and $-1$ for $Q$. 
The Wigner function is an especially important case because,
although it is not positive-definite, integrals of the Wigner function yield the marginal distributions
for position and momentum. The position marginal distribution is thus
\begin{equation}
\label{eq:marginal}
	P(q_\theta)=\int_{-\infty}^\infty W(q_\theta,q_{\theta+\pi/2})dq_\theta.
\end{equation}

Titulaer and Glauber introduced ``generalized coherent states'' as states that are fully
coherent with respect to the coherence functions but are not eigenstates of the
annihilation operator~$\hat{a}$~\cite{TG65}. These states have Poisson number distributions
but allow an arbitrary phase relationship between coefficients in the Fock representation
of the state.
For $\not\! \text{d}\varphi:=\text{d}\varphi/2\pi$
\begin{align}
\label{eq:gencohstates}
	\left|\sqrt{\bar{n}};\bm{\vartheta}\right\rangle
		&= e^{-\bar{n}/2}\sum_{n=0}^\infty \frac{\bar{n}^{n/2}}{\sqrt{n!}}e^{i\vartheta_n}|n\rangle
			\nonumber	\\	&=
		\int_0^{2\pi} \not\!\text{d}\varphi \left|\sqrt{\bar{n}}e^{i\varphi}\right\rangle 
		\sum_{n=0}^\infty e^{i(\theta_n-n\varphi)},
\end{align}
with the last line corresponding to the representation of the generalized
coherent state as a superposition of coherent states on a circle, and this representation 
on a circle was introduced by Bia\l ynicki-Birula~\cite{Bia68}.

Spiridinov~\cite{Spi95} showed that these generalized coherent state are eigenstates of 
a generalized annihilation operator that holds the number operator~$\hat{n}$ invariant.
Physically Spiridonov's transformation corresponds to a number-sensitive rotation; optically we can understand this transformation
as a generalization of the ideal single-mode optical Kerr nonlinearity, which effects a
phase shift that is a function of field strength, or, equivalently, photon number for the single-mode
field.

\subsection{Superpositions of coherent states}
\label{subsec:superpositions}

Whereas the coherent state is regarded as the closest quantum optical description to
a classical coherent field, superpositions of coherent states exemplify the strangeness of
quantum theory. In general any pure state of the field~$\left|\psi\right\rangle$ can be written as a 
superposition of coherent states according to the expression
\begin{equation}
	\left|\psi\right\rangle = \int \frac{d^2\alpha}{\pi} \langle\alpha\left|\psi\right\rangle\,\left|\alpha\right\rangle.
\end{equation}
As the coherent states form an overcomplete basis, it is not surprising that every state
can be expressed as a superposition of coherent states.

Interestingly, the overcompleteness of the coherent-state basis
allows quite different ways to write the superposition. One particularly important case is
the superposition of coherent states on the circle, which we have encountered
in Subsec.~\ref{subsec:quoptics} in studying the Titulaer-Glauber
generalized coherent states~\cite{TG65}. Other states, can also be expressed in
this way. For example, the Fock number state
has the appealing representation~\cite{Bia68,BK91,BK95}
\begin{equation}
\label{eq:numbercircle}
	|n\rangle = \frac{1}{\sqrt{\Pi_n(m)}}\int\not\! d\varphi e^{-im\varphi}
		\left|\sqrt{m}e^{i\varphi}\right\rangle~,
\end{equation}
with $\Pi$ the Poisson distribution~(\ref{eq:Poisson}) and $\not\!d:=d/2\pi$.
Expression~(\ref{eq:numbercircle}) is a superposition over coherent states 
with complex amplitude restricted to having modulus $\sqrt{m}$. States can also be 
expressed as superpositions of coherent states on lines or other subspaces of
the $\alpha$ parameter space.

Evolution of a coherent state under an ideal optical Kerr nonlinearity
\begin{equation}
\label{eq:idealKerr}
	\vartheta(\hat{n})=\omega\hat{n}+\lambda\omega^2\hat{n}^2
\end{equation}
yields a particular form of generalized coherent state~\cite{Mil85,MH86,DM89}.
Yurke and Stoler~\cite{YS86,YS87} showed that a superposition of two
coherent states could be obtained under this evolution,
and its generalization to $\vartheta(\hat{n})\propto\hat{n}^k$ could
be expressed as a finite superposition of coherent states with different
phases for certain evolution times.
In fact the Titulaer-Glauber generalized coherent~\cite{TG65} can
be expressed as a superposition of a finite number of coherent states
on the circle  for $\vartheta_{n+N}=\vartheta_n$ for some $N$ and for all~$n$~\cite{Bia68,Sto71}.

Spirodonov~\cite{Spi95} identified two other interesting cases of generalized
coherent states: $q$-deformed coherent states for which $\vartheta_{n+N}=q\vartheta_n$
and parity coherent states $\vartheta_n=n\pi$. The parity operator is $\exp\{i\pi\hat{n}\}$,
and the (unnormalized) parity coherent state is given by
\begin{equation}
\label{eq:balancedcat}
	\text{e}^{-i\pi/4}\left|\alpha\right\rangle+\text{e}^{i\pi/4}|-\alpha\rangle,
\end{equation}
which is a special case of the superpositions of coherent states with equal complex
field amplitudes and equal phase separations studied by Bia\l ynicki-Birula~\cite{Bia68}
and Stoler~\cite{Sto71}. These superpositions are discussed in more detail in 
Subsec.~\ref{subsec:superpositions}.

As we saw in Subsec.~\ref{subsec:quoptics}, superpositions of coherent states on a circle
can arise via evolution of a coherent state according to a generalized Kerr nonlinearity,
yielding an evolution operator~$\exp\{i\vartheta(\hat{n})\}$. The equally weighted
superposition of two coherent states that are $\pi$ out of phase with each 
other~(\ref{eq:balancedcat}), introduced by Yurke and Stoler~\cite{YS86},
has been termed a ``Schr\"{o}dinger cat state'', or ``cat state'' for short, because the coherent state
is regarded as being an essentially classical field state, and the superposition
of two highly distinct coherent states is reminiscent of the Schr\"{o}dinger's 
cat being described as being in the state $|\text{`live'}\rangle+|\text{`dead'}\rangle$.

The term Schr\"{o}dinger cat state has also been applied to the `even' and `odd' 
coherent states~\cite{DMM74},
\begin{equation}
	\left|\alpha\right\rangle_\pm=N_\pm\left(\left|\alpha\right\rangle\pm|-\alpha\rangle\right)
\end{equation}
for
\begin{equation}
	N_+=\frac{\exp\left(|\alpha|^2\right)}{2\sqrt{\cosh|\alpha|^2}},\,
	N_-=\frac{\exp\left(|\alpha|^2\right)}{2\sqrt{\sinh|\alpha|^2}}.
\end{equation}
The even-odd terminology refers
to the fact that the photon number distribution is non-zero only for even photon
number in the case of the even coherent state~$\left|\alpha\right\rangle_+$ and is non-zero
only for odd photon number in the case of the odd coherent state~$\left|\alpha\right\rangle_-$.
As this state does not have a Poisson number distribution, it cannot be evolved via
a generalized unitary Kerr evolution from the coherent state 
but is a Titulaer-Glauber generalized coherent state~\cite{TG65}
for which $\vartheta(\hat{n}+2\openone)=\vartheta(\hat{n})$.
Even and odd coherent states may arise by a conditional Jaynes-Cummings 
evolution~\cite{Gea90,DMM+93}.

Detection of Schr\"{o}dinger cat states may be achieved by optical homodyne 
detection, with the measurement results converging to the marginal distributions for
canonical position and momentum~\cite{YS86,YS87} even in the presence of decoherence~\cite{MT87,TM87}.
Let us consider the `balanced cat' of Eq.~(\ref{eq:balancedcat})
with its equally weighted superposition of two coherent states $\pi$ out of phase. 
If the local oscillator is in-phase with either of the coherent states, the marginal
distribution is equivalent to that for an incoherent mixture of such coherent states. The 
marginal distribution for the conjugate quadrature exhibits interference fringes that 
yield information on how coherent, or pure, the superposition state is.

So far we have considered superpositions of single-mode coherent states,
but a superposition of multimode coherent states of the type~(\ref{eq:prodcohstate})
is also allowed. Such a state can be written as
\begin{equation}
\label{eq:hwnecs}
	\int d\mu(\bm{\alpha}) f(\bm{\alpha})|\bm{\alpha}\rangle
\end{equation}
for which the measure $d\mu(\bm{\alpha})$ can be over the entire parameter space
or over subspaces for which the set $\{\bm{\alpha})\}$ is an overcomplete basis.
Before proceeding onto studies of this superposition of multimode coherent states,
we consider how coherent states and their superpositions are generalized to
Lie groups and algebras other than the Heisenberg-Weyl group for simple harmonic oscillators.

\subsection{Lie coherent states and their superpositions}
\label{subsec:supgen}

The term  ``generalized coherent state'' has been used in Subsec.~\ref{subsec:superpositions}
to refer to loosening the phase relation between elements of the coherent state expressed
as a superposition in the Fock basis; we have been careful to refer to these as
`Titulaer-Glauber generalized coherent states''~\cite{TG65}.
The term ``generalized coherent state'' has also been 
applied to establishing coherent states for general Lie groups. 
Here we refer to Lie group and algebra generalizations of coherent states as ``Lie coherent states''.
Where the specific Lie algebra is specified, the notation for the algebra replaces ``Lie'',
e.g.\ ``$\mathfrak{su}(2)$ coherent state''.
For $n$ simple harmonic oscillators, the operator algebra is $\mathfrak{hw}$($n$)
given in~Eq.~(\ref{eq:hwn}), and the Lie coherent state for $\mathfrak{hw}$($n$)
is the multimode product coherent state~(\ref{eq:prodcohstate}).

Whereas $\mathfrak{hw}$($N$) coherent states are 
(i)~displaced vacuum states (orbits of the vacuum state
under the action of the displacement operator~$D(\bm{\alpha})$, 
(ii)~eigenstates of the lowering operator~$\hat{\bm{a}}$, and 
(iii)~minimum-uncertainty states, 
some sacrifice must be made in defining coherent states for other Lie groups.
A basis set of operators for a Lie algebra can be expressed as lowering operators
analogous to $\hat{a}$, the conjugate raising operators, and the Cartan subalgebra,
which is a set of mutually commuting elements of the algebra.

A Lie algebra generates a $k$-parameter Lie group with the dimension of the Cartan subalgebra being~$k$.
Studies of superpositions and entanglement of coherent states have so far focused
primarily on one-parameter Lie groups (with the exception of one study on $\mathfrak{su}$(3) coherent states~\cite{NS01}),
so we restrict our attention to that case; in fact
we can concentrate on $\mathfrak{su}(2)$ and $\mathfrak{su}(1,1)$.

Entanglement of $\mathfrak{su}(2)$ and $\mathfrak{su}(1,1)$ coherent states were studied by
Wang and Sanders~\cite{WSP00}. 
The corresponding algebras are
\begin{equation}
\label{eq:su2algebra}
	[\hat{J}_+,\hat{J}_-]=2\hat{J}_z, [\hat{J}_z,\hat{J}_\pm]=\pm\hat{J}_\pm
\end{equation}
and
\begin{equation}
\label{eq:su11algebra}
	[\hat{K}_+,\hat{K}_-]=-2\hat{K}_z, [\hat{K}_z,\hat{K}_\pm]=\pm\hat{K}_\pm,
\end{equation}
respectively, for $\mathfrak{su}(2)$ and $\mathfrak{su}(1,1)$,
with $J$ used for the compact SU(2) group and $K$ used for the non-compact SU(1,1) group.
The Cartan subalgebras are $J_z$ for $\mathfrak{su}(2)$ and $K_z$ for $\mathfrak{su}(1,1)$.
The Casimir
invariants are $\hat{J}^2$ with spectrum $j(j+1)$ for irreducible representation, or irreps, 
indexed by $j\in\{0,1/2,1,3/2,\ldots\}$,
and $\hat{K}^2$ with spectrum $k(k-1)$ for irreps indexed by
\begin{equation}
	k\in\{1/2,1/3/2,2,\ldots\}.
\end{equation}
Within a given irrep, an orthonormal basis is given by
\begin{equation}
	\{|j\, m\rangle;m=-j,j+1,\ldots,j\},
\end{equation}
such that
\begin{equation}
	\hat{J}_-|j\, m\rangle=\sqrt{j-m+1}|j\, m-1\rangle,\,|j\, -j-1\rangle\equiv 0
\end{equation}
with $\{|k\, n\rangle;n=0,1,2,\ldots\}$ for $\mathfrak{su}(2)$.
The $\mathfrak{su}(1,1)$ can be constructed in a similar way.

For non-compact groups, there are two inequivalent ways to construct coherent states:
(i)~as eigenstates of the lowering operator and (ii)~as orbits of a minimum uncertainty
state, analogous to the orbit of the vacuum state under the displacement operator
(\ref{eq:displacement}). Highest- or lowest weight states (states that are annihilated
by the raising and lowering operators, respectively) are typical minimum-uncertainty states for
the two groups under consideration.

In 1971, Barut and Girardello~\cite{BG71} introduced ``new coherent states'' for non-compact
groups based on criterion~(ii).
They identified the lowering operator(s) and found eigenstates for this operator. For $\mathfrak{su}(1,1)$,
the lowering operator is $\hat{K}_-$, and the Barut-Girardello $\mathfrak{su}(1,1)$ coherent state is 
\begin{equation}
\vert k \, \eta \rangle_\text{BG}=\sqrt{\frac{|\eta |^{k-1/2}}{I_{2k-1}(2|\eta |)}}
\sum_{n=0}^\infty \frac{\eta ^n}{\sqrt{n!\Gamma (n+2k)}}\vert k \, n\rangle ,
\end{equation}
which satisfies $\hat{K}_-|k \,\eta\rangle_\text{BG}=\eta |k\,\eta\rangle_\text{BG}$,
for $I_\nu (x)$ the modified Bessel function of the first kind.

For both non-compact and compact groups, coherent states can be defined analogously
to the displaced vacuum state by introducing a minimum-uncertainty state such as 
the highest- or lowest-weight state in the representation and considering a generalization
of the displacement operator. For $\mathfrak{su}(2)$, the usual minimum-uncertainty state is 
the highest-weight state $|j\, j\rangle$ although the lowest-weight state is used as well.

For $\mathfrak{su}(1,1)$, the lowest-weight state $|k\, 0\rangle$ is used.
For SU(2), the analogue to the displacement operator is the ``rotation operator''
\begin{equation}
\label{eq:Rotation}
	R(\theta,\varphi)=\left\{\frac{\theta}{2}\left[
		e^{-i\varphi}\hat{J}_--e^{i\varphi}\hat{J}_+\right]\right\},
\end{equation}
and, for SU(1,1), the analogue is the ``squeeze operator''~\cite{YMK86}
\begin{equation}
\label{eq:Squeeze}
	S(\xi)=\exp\left\{\xi\hat{K}_+-\xi^*\hat{K}_-\right\}.
\end{equation}
The term `rotation operator' applies for SO(3), which is the rotation group in three-dimensional
Euclidean space, and the term has been extended to apply to SU(2), which is a double
covering group of SO(3). The term `squeeze operator' is used here because two-boson realizations of $\mathfrak{su}(1,1)$ are
\begin{equation}
\label{eq:Kboson}
	\hat{K}_-=\hat{a}^2 \, (k=1/4),\;  \hat{K}_-=\hat{a}\hat{b}\,(k=3/4),
\end{equation}
and either of these realizations of $\mathfrak{su}(1,1)$ converts the unitary operator~(\ref{eq:Squeeze})
to the usual one-mode and two-mode squeeze operators in quantum optics for
$k=1/4$ and $k=3/4$, respectively.

The $\mathfrak{su}(2)$ coherent states were first introduced as ``atomic coherent states''~\cite{ACGT72}.
These states are given by 
\begin{equation}
	|j;\theta,\phi\rangle=R(\theta,\phi)|j\, j\rangle
\end{equation}
and form an overcomplete basis of the Hilbert space~\cite{Gil72,Per72}.
The coherent states are orbits of the minimum-uncertainty state under the action of
a group element. Perelomov undertook a general analysis of such coherent states
for any Lie group, and these Lie coherent states are known as Perelomov coherent
states~\cite{Per86}.

We refer to the Lie coherent state using the notation~$|\ell\,\xi\rangle$
with $\ell$ the irrep parameter (not required for the $\mathfrak{hw}$($n$) algebra)
and $\xi$ the orbit parameter. This notation applies equally to eigenstates of
the lowering operator (as for the Barut-Girardello states) and for orbits
of minimum-uncertainty states (as for the Perelomov states).
The multipartite Lie coherent state is designated by $|\ell \,\bm{\xi}\rangle$,
which is a product of Lie coherent states $|\ell\, \xi_i\rangle$ all from the same
irrep.

So far we have only considered entangled coherent states where each party in the state has the same coherent-state structure.
For example the entangled coherent state can be a superposition of a tensor product of $\mathfrak{hw}$(1) coherent states
or a superposition of tensor product of $\mathfrak{su}$(2) coherent states or so on.
On the other hand a `hybrid' entangled coherent state could be constructed as a superposition of tensor
products of coherent states with coherent states in the tensor-product space corresponding to different types of coherent states.
Such hybrid coherent states have not been studied but have been realized experimentally in a limited way:
hybrid $\mathfrak{hw}(1)$-$\mathfrak{su}(2)$ entangled coherent state
are realized in cavity quantum electrodynamics experiments as entangled atom-field states~\cite{DMM+93,DBRH96,BHD+96}.

\subsection{Entangled coherent states}
\label{sec:ecs}

A superposition of mutimode or mutipartite coherent states can be expressed in general as~\cite{WSP00}
\begin{equation}
\label{eq:generalecs}
	\int d\mu_\ell(\bm{\xi})f_\ell(\bm{\xi})|\ell\bm{\xi}\rangle
\end{equation}
with the state~$|\ell\bm{\xi}\rangle$ the Lie coherent state.
For the usual case of harmonic oscillators,
corresponding to the algebra $\mathfrak{hw}$($n$), the index $\ell$ is superfluous, and we let
$\bm{\xi}$ be replaced by $\bm{\alpha}$ to obtain~(\ref{eq:hwnecs}).
This superposition is not entangled if there
exists any representation for the pure state such that the state can be expressed
as a product state over the modes. Otherwise the state is entangled. Thus entangled
coherent states are a special case of superpositions of multimode coherent states,
but a rather large and especially interesting class of states.

The entangled state~(\ref{eq:generalecs}), which is expressed as an integral
of product coherent states, can be reduced to a sum if the function $f_\ell(\bm{\xi})$
can be expressed as a sum of delta functions
\begin{equation}
	f_\ell(\bm{\xi})=\sum_i f_\ell(\bm{\xi}_i)\delta(\xi-\xi_i).
\end{equation}
Then 
\begin{equation}
\label{eq:discreteecs}
	\int d\mu_\ell(\bm{\xi})f_\ell(\bm{\xi})|\ell\bm{\xi}\rangle
		= \sum_i f_\ell(\bm{\xi}_i)|\ell\bm{\xi}_i\rangle~.
\end{equation}
In the single-particle (equivalently single-mode) case, such states are the Titulaer-Glauber coherent states~\cite{TG65}.

As an interesting example of discrete bipartite entangled coherent states,
van Enk studied the discrete ``multidimensional entangled coherent states''~\cite{vEn03}
\begin{equation}
\label{eq:Phi_M}
|\Phi_M\rangle=\frac{1}{\sqrt{M}} \sum_{q=0}^{M-1} e^{i\phi_q}
	\left| \alpha e^{2\pi iq/M},\alpha e^{2\pi iq/M}\right\rangle,
\end{equation}
which are generated by an ideal nonlinear Kerr evolution~(\ref{eq:idealKerr}),
to characterize the entangling power.
He shows that such states have infinite entanglement for short times~$\tau$ infinitesimally short evolution times and finite entanglement after finite times.
Finite discrete superpositions can serve as a resource for quantum teleportation.
Showed that, for very small losses for multidimensional entangled coherent states, approximately 2.89 ebits are lost per absorbed photon,
which could be useful for creating entangled coherent states with a fixed amount of entanglement~\cite{vEn05}.

Entangled coherent states overlap conceptually with the pair coherent state,
which was introduced as the joint eigenstate of the two-mode, annihilation operator~$\hat{a}_1\hat{a}_2$
and the number difference operator $\hat{n}_1-\hat{n}_2$~\cite{Aga86,Aga88,Lee98}.
Defining the pair coherent state by $|\zeta,q\rangle$, with $\zeta$ the eigenvalue of
the pair annihilation operator and $q$ the eigenvalue of the photon number difference operator.
These states exhibit sub-Poissonian statistics, correlated number fluctuations, squeezing, and violations of photon Cauchy-Schwarz inequalities.

Pair coherent states are an example of $\mathfrak{su}(1,1)$ coherent states, represented as two-boson realizations.
The pair annihilation operator can be expressed according to
the algebra~(\ref{eq:su11algebra}): the pair annihilation operator is $\hat{K}_-$ and
the photon number difference operator is $\hat{K}_z$.

The pair coherent state is an example of an entangled coherent state, which is evident
by expressing the pair coherent state as
\begin{equation}
\label{eq:paircoherentstateasecs}
	|\zeta,q\rangle = \int_0^{2\pi} \not\! d\varphi \frac{N_q e^{|\zeta)}}
		{\left[\sqrt{\zeta}e^{i\varphi}\right]^q}
		|\sqrt{\zeta}e^{i\varphi}\rangle|\sqrt{\zeta}e^{-i\varphi}\rangle
\end{equation}
with $q$-dependent normalization constant
\begin{equation}
	N_q = \frac{|\zeta|^{q/2}}{\sqrt{I_q(2|\zeta|)}},
\end{equation}
for $I_q$ the modified Bessel function of the first kind,
as expressed by Gerry and Grobe~\cite{GG95}.

The Schr\"{o}dinger cat state concept, which corresponds to a superposition of coherent
states, was extended to a superposition of pair coherent states by 
Gerry and Grobe~\cite{GG95}. Specifically the superposition of two pair coherent states
can be expressed as
\begin{equation}
	|\zeta,q,\phi\rangle=\frac{|\zeta,q\rangle+e^{i\phi}|-\zeta,q\rangle}
		{\sqrt{2+2N_q^2\cos\phi\sum_{n=0}^\infty \frac{(-1)^n|\zeta|^{2n}}
		{n!(n+q)!}}}
\end{equation}
which is an eigenstate of the squared pair annihilation operator with eigenvalue $|\zeta|^2$.
This superposition of pair coherent states is also an entangled coherent state, which is
a superposition of two entangled coherent states of the type~(\ref{eq:paircoherentstateasecs}).

\section{Implementations}
\label{sec:implementations}

Entangled coherent states are fragile due to their entanglement but are nonetheless implementable if the conditions are right.
Furthermore the states serve as a resource for quantum information processing so they have utility and value hence are worth making.
Many theoretical proposals exist for constructing entangled coherent states in the laboratory but so far the paramount experimental 
demonstration uses a photon-subtraction technique on two approximate Schr\"{o}dinger cat states
so that the source of the photon is indeterminate~\cite{OFTG09,SSG+10}.

\subsection{Parametric amplification and photodetection}

Consider two physically separated states of light of the type $\left|\alpha\right\rangle+\left|-\alpha\right\rangle$. 
These two fields are each passed through a separate beam splitter such that each field loses a small fraction of its energy.
The extracted part of each field is brought together after using a phase shifter to impose a $\phi$ phase difference between the two beams.
The fields are combined at a beam splitters with a photon counter at one output port.
The effect of this final beam splitter is to ensure that the detected photon is equally likely to have come from either beam.
The resultant two-mode state conditioned on registering a single photon count is
\begin{align}
\label{eq:ecskitten}
	-i\sin\frac{\phi}{2}&\left|\alpha\right\rangle\left|\alpha\right\rangle
		-\cos\frac{\phi}{2}\left|-\alpha\right\rangle\left|\alpha\right\rangle
			\nonumber	\\
		&+\cos\frac{\phi}{2}\left|\alpha\right\rangle\left|-\alpha\right\rangle
		+i\sin\frac{\phi}{2}\left|-\alpha\right\rangle\left|-\alpha\right\rangle.
\end{align}
This is the concept behind the successful experimental creation of a close approximation of this state,
and the success of the process is verified by optical homodyne tomography on the resultant state~\cite{OFTG09}.

The actual experiment involves using a pulsed optical parametric amplifier as a source of squeezed vacua. 
The cat state with small amplitude~$\alpha$,
known as a ``kitten state'',
can be prepared by subtracting a single photon~\cite{OTLG06,ODTG07}.
Using this principle, the two output modes squeezed vacua of orthogonal polarizations are recombined at a polarizing beam splitter.
A small fraction of each field goes to the photon counter, which conditions the rest of the field going out the other beam splitter port into 
the entangled coherent state~(\ref{eq:ecskitten}).
The state is then tomographically characterized.

\subsection{Nonlinear optics}
\label{subsec:optics}

The earliest proposals for creating entangled coherent states were expressed in the context of quantum optical fields
interacting via a third-order optical nonlinearity known as a Kerr nonlinearity~\cite{Wei08}.
The optical Kerr nonlinearity features a refractive index $n_0+n_2I$,
which is the sum of a linear refractive index~$n_0$ and a second term that is proportional to the field strength typically characterized by `intensity'~$I$.
The Kerr effect, as a third-order optical nonlinearity,
is a special case of four-wave mixing.

The term `cross-Kerr nonlinearity' is ubiquitous in the entangled coherent state literature and refers to the phenomenon that one field experiences
a phase shift component that is proportional to the strength of the other field.
The cross-Kerr effect thus leads to `cross-phase modulation', which is specifically the phase shift of one field due to the intensity of the other.
In the quantum analysis, the phase shift depends on the photon number in the other beam.
Of course the phase shift of the beam also depends on its own strength,
and this is known as `self-phase modulation'.

The first proposal for generating entangled coherent states was introduced by Yurke and Stoler in 1986~\cite{YS86,YS87}
followed by the work of Mecozzi and Tombesi in 1987\cite{MT87,TM87} and others~\cite{SR99,SR00}.
The entangled coherent state became the chief object of interest in work by Sanders a few years later
with a proposed implementation that inserted a Kerr nonlinearity into one path of a Mach-Zehnder interferometer
(often called a `nonlinear interferometer' for short)~\cite{San92,San92E}
and has been the subject of further study~\cite{Ger99,RS98,SR99,SR00}.
Related to this approach, if an appropriate superposition of two coherent states is provided,
then a beam splitter transformation alone suffices to produce an entangled coherent state from this resource~\cite{AAKJ04}

The use of a Kerr nonlinearity to entangle the coherent state with a vacuum state,
which is a special case of bipartite entangled coherent states,
was generalized by Luis to show how to entangle any state with the vacuum state~\cite{Luis02}.
Wang showed how a nonlinearity coupled with linear optical elements
can be employed to generate general bipartite entangled non-orthogonal states~\cite{Wan02}.

Variants of nonlinear interacting propagating field realizations of entangled coherent states have been studied.
Slow light in a medium with double electromagnetically induced transparency could be used to enable entangled coherent state generation~\cite{PKH03,GK05}.
Entanglement could first be prepared in matter qubits then transferred to fields to make entangled coherent states by exploiting a cross-Kerr nonlinearity~\cite{LPO+06}.
Nonlocal preparation of a bipartite entangled coherent state, 
where `nonlocal' means that the two fields being entangled never meet or directly interact,
could be produced by sending a photon through a Mach-Zehnder interferometer with a nonlinear Kerr medium in each of its two paths,
and separate coherent states are sent through each of these two nonlinear media~\cite{GG07}. 
The bipartite entangled coherent state is post-selected by detecting from which port the photon leaves:
whichever port the photon leaves from post-selects the nonlocal two-mode field in one of two entangled coherent states.

Generation of various exotic forms of entangled coherent states have been investigated.
Greenberger-Horne-Zeilinger and W~types of entangled coherent states could be produced
with propagating fields using linear optics and Kerr nonlinearities~\cite{JA06,LY09}.
Similarly cluster-type entangled coherent state can be generated with a nonlinear medium and a laser driving field~\cite{WSL08,AKK11,HDF+11}.

\subsection{Cavity quantum electrodynamics}

Entangled coherent states can be created in cavity fields rather than in propagating fields,
which has the advantage of large effective nonlinearities.
The nonlinearity in the medium could be a macroscopic optical Kerr medium or one or more multilevel atoms.
For example a multimode entangled coherent state can be prepared by letting a
a single atom traverse two or more single-mode cavities, each occupied initially by a coherent state,
and then post-selecting on the atomic state~\cite{DMM+93,Zhe98,KA99,SZG02,SZG02E,BCAB08}.
An unbalanced (i.e.\ unequally weighted) entangled coherent state could be produced in a double-cavity system~\cite{WS93}.
Alternatively just one cavity that supports a multimode field is an alternative to multiple-cavity generation of entangled coherent states~\cite{GZ97,KA99,SAW03,Guo04,MMZ07}.

Matter-wave interferometry could assist in preparing entangled coherent states.
If a two-mode cavity can be prepared in a pair coherent state,
then this state can be transformed into an entangled coherent state by the following procedure.
Atoms are sent through a double slit and then interact with the two-mode cavity field.
Atomic-position detection subsequently post-selects the two-mode field into an entangled coherent state~\cite{ZG97}.

Artificial atoms, such as quantum dots~\cite{WFS03} or Cooper-pair boxes~\cite{BA06},
can replace real atoms to produce entangled coherent states in cavity quantum electrodynamics.
The microwave regime could prove to be quite appropriate for generating entangled coherent states with Rydberg atoms in 
millimeter-wave superconducting cavities~\cite{MAY+05,HBR07}.

As with propagating fields interacting with a Kerr medium,
exotic entangled coherent states such as Greenberger-Horne-Zeilinger, W~\cite{CM07} and cluster-type entangled coherent states,
can also be created in cavities~\cite{BCAB08,LY09,Tan09,WW10,Tan10}.
Modified entangled coherent states,
such as the ``single-mode excited entangled coherent states'',
could also be created in a cavity quantum electrodynamics setting~\cite{XK06}.

\subsection{Motional degrees of freedom}
\label{subsec:ion}

Instead of creating entangled coherent states in electromagnetic field modes,
motional degrees of freedom can be used instead.
Vibrational degrees of freedom for a single trapped ion in two dimensions~\cite{Ger97} or of two trapped ions~\cite{Zhe01,WS01,LWJ03}
or for collective modes (e.g.\ center-of-mass or breathing modes) of many trapped ions~\cite{SdMZ02, Zhe05},
can be transformed into entangled coherent states.
Ion traps can be combined with cavity quantum electrodynamics set-ups to make hybrid entangled coherent states between
electromagnetic and motional degrees of freedom~\cite{ZPM02,LGZK06}.

As an example of creating an entangled coherent state in the vibrational degrees of freedom of a single trapped ion in two dimensions,
consider the interaction Hamiltonian~\cite{Ger97}
\begin{equation}
	\hat{H}_\text{Int}
		=-\hbar\chi\left(\hat{a}^\dagger\hat{a}-\hat{b}^\dagger\hat{b}\right)
			\left(\hat{\sigma}_++\hat{\sigma}_-\right)
\end{equation}
with $\hat{a}$ and $\hat{b}$ the annihilation operators for each of the two vibrational modes,
$\chi$ a coupling coefficient and $\hat{\sigma}_+=\hat{\sigma}_-^\dagger$ the electronic-energy lowering operator.
Both vibrational modes are initialized in coherent states $|\alpha\rangle$ and~$|\beta\rangle$
and the ion in the ground state~$|\text{g}\rangle$.
At time~$t$ the combined (unnormalized) state for the atom and the two vibrational states is
\begin{align}
	|\text{g}\rangle&\left(\left|\alpha\text{e}^{i\chi t},\beta\text{e}^{i\chi t}\right\rangle
		+\left|\alpha\text{e}^{-i\chi t},\beta\text{e}^{-i\chi t}\right\rangle\right)
			\nonumber	\\
	&+|\text{e}\rangle\left(\left|\alpha\text{e}^{i\chi t},\beta\text{e}^{i\chi t}\right\rangle
		-\left|\alpha\text{e}^{-i\chi t},\beta\text{e}^{-i\chi t}\right\rangle\right)
\end{align}
for $|\text{e}\rangle$ the excited state of the atom.
The entangled coherent state can be created post-selectively by measuring the electronic state of the ion.

Ion traps could be used to create multipartite entangled coherent states using entanglement swapping operations~\cite{WS01}.
Consider two identical ions with each initially prepared in a superposition of ground and excited state and the
center-of-mass and breathing modes each initially prepared in coherent states with the same amplitude and phase.
Then two distinct Raman beams are directed at the two ions independently.
One beam is directed at the first ion in order to couple it to the fundamental mode,
and the second beam is directed at the second ion in order to couple it to the breathing mode.

This selective coupling of ion electron levels to motional modes is achieved by choosing judicious Raman parameters.
Subsequently Bell-state measurements of the two-ion electronic states result in the two motional modes `collapsing',
or being post-selected, into entangled coherent states. 
This principle is readily extended to the multimode entangled coherent state case by extending from two to 
as many ions as desired naturally accompanied by as many vibrational modes.
The multi-ion electronic state is projected onto a maximally entangled state (generalized Bell measurement)
thereby resulting in the vibrational modes being in a multimode (or `multipartite' entangled coherent state)~\cite{WS01}.

Other physical realizations that are amenable to creating entangled coherent states in motional
degrees of freedom include nano-cantilevers~\cite{BA06} and movable nano-mirrors~\cite{BJK97,Zhe00}.
More pointedly, entangled coherent states can in principle be realized in any system
that can be described as harmonic oscillators with appropriate nonlinear coupling and sufficiently low loss and decoherence.

\subsection{Bose-Einstein condensates}
\label{subsec:BEC}

Bose-Einstein condensates have an inherently high nonlinearity due to atomic collision terms, and it is possible to prepare
two separate Bose-Einstein condensates of three-level atoms
(corresponding to different electronic states of the same atoms)
into phase-locked coherent states and couple them together via a Raman interaction~\cite{JGS+01}.
This approach could be used to entangle an arbitrary state of one Bose-Einstein condensate with a ground state of the other Bose-Einstein condensate~\cite{Luis02}.
Alternatively entangled coherent states could be generated in Raman-coupled Bose-Einstein condensates~\cite{KZK04}.

Nonlocal preparation of distant entangled coherent states could be possible using electromagnetically induced transparency~\cite{KCP09}. 
In this scheme, two strong coupling laser beams and two entangled probe laser beams prepare two distant Bose-Einstein condensates in 
electromagnetically induced transparency coherent population states then forced to interact.
The two Bose-Einstein condensates are initially in a product coherent state while the probe lasers are initially entangled. 
The final preparation step involves performing projective measurements upon the two outgoing probe lasers.

\section{Nonclassical properties}
\label{sec:nonclassical}

Entangled coherent states are highly nonclassical states but are peculiar in that they are expressed as an entanglement of the most classically
well behaved states we know: coherent states.
Thus entangled coherent states are especially intriguing in studies of nonclassicality because the state represents an entanglement of classically meaningful
descriptions of objects.

Nonclassicality is studied through a variety of measures including squeezing~\cite{Cha92,SR00,ZG11}, 
sub-Poissonian photon statistics~\cite{Cha92},
violations of Cauchy-Schwarz inequalities~\cite{Cha92},
complementarity between particle-like and wave-like features of entangled coherent states~\cite{RS98},
violations of a Bell inequality~\cite{San92,San92E,MSM95, GDR99,MAY+05,HBR07,GMB09} or Leggett's inequality~\cite{LPJ11} for testing local realism,
and entanglement properties such as index of correlation~\cite{WSP00}, entanglement of formation~\cite{LX03} and other measures~\cite{ZG11}.
Nonclassicality of generalized entangled coherent states,
such as $\mathfrak{su}$(2) and $\mathfrak{su}$(1,1) states~\cite{WSP00,GBHA08}
and photon-added entangled coherent states~\cite{ZX09} has been studied as well.

\subsection{Complementarity}
\label{subsec:complementarity}

Complementarity in double-slit~\cite{WZ79} and two-channel interferometery~\cite{SM89}
studies is well understand for single-particle inputs.
Single photons exit wholly from either one or the other port of a beam splitter
but exhibit strong fringe visibility when the experiment is modified by replacing the beam splitter by an interferometer~\cite{GPY88}.
Complementarity may be understood by thinking of the photon's path state as entangled:
a superposition of the photon traversing one path (e.g.\ through one slit or down one channel of the interferometer) and a vacuum state in the other path
and the reverse case.
Now consider, instead of a photon in one path and vacuum in the other,
we have a coherent state in one path and a vacuum in the other. 
Would complementarity be manifested and observable in that case?

Rice and Sanders showed that, in principle, a form of complementarity is present,
but the notion of a phase shifter, which is a simple linear optical element for a photon,
is complicated for a coherent state yet necessary to observe the undularity of the coherent state
in the context of entangled coherent states~\cite{RS98}.
Joint photodetection at the two interferometer output ports~\cite{WS93} can reveal anticorrelation of the nonlinear Mach-Zehnder interferometer output
thereby revealing `corpuscularity' of the coherent state
analogous to the anticorrelation revealing corpuscularity for a single photon~\cite{RS98}.
The coherent state is thus `seen' to follow one path or another and not be split.

The ideal coherent state phase shifter would correspond to the unitary transformation $\exp(-\text{i}\phi|\alpha\rangle\langle\alpha|)$
for imposed phase shift~$\phi$ and could be created in approximate form in a highly nonlinear medium with appropriate parameters~\cite{RJS01}.
The creation of this phase shifter would enable other types of tests of complementarity such as performing
two-coherent-state interferometry, even with large numbers of photons.
Two-coherent-state interferometry is analogous to two-particle quantum interferometry but with the single-particle Fock state replaced by a coherent state~\cite{RJS01}.

\subsection{Entanglement}
\label{subsec:entanglement}

The nomenclature ``entangled coherent state'' demands quantification of the degree of entanglement of such states. 
There is more than one way to study entanglement of such states. 
One can consider Bell inequalities or generalizations thereof,
perhaps to test local realism or just to show non-factorizability.
Another approach to studying entanglement of these states is to recognize that unentangled coherent states
have Gaussian statistics and then use the covariance properties to quantify the degree of entanglement
in such states~\cite{DdCM02}. An alternative approach considers the entangling power of operations
that produce entangled coherent states~\cite{vEn03}. Each of these approaches is 
challenging because the Hilbert spaces are infinite-dimensional and the entanglement is between non-orthogonal states~\cite{MSM95,Wan02}.

Quantifying entanglement can instead by studied in the context of performing a quantum information processing task
such as quantum teleportation~\cite{BBC+93}.
Teleportation enables a qubit to be sent from one party to another through a classical channel
by sending instead two bits of information and consuming one ``ebit'', or entangled bit (two maximally entangled qubits) of a prior shared entanglement resource.

Entanglement can then be quantified by determining how well entangled coherent states
serve as the `quantum channel' (i.e.\ the consumable prior shared resource) 
for teleporting another state.
This other state could be a qubit corresponding to a superposition of single-mode coherent states (`cat state')~\cite{vEH01,JK02,JBK+02,An03,JAG+03,AAKJ04,MRVS10}.
The entangled coherent state can supply an entire ebit of resource despite being an entanglement of non-orthogonal states~\cite{Wan01,vEH01,JKL01}.
An alternative approach to studying quantum resources considers how well a given a resource serves to teleport
all or part of an entangled coherent state~\cite{Wan01,JBS02,CGNJ06,LK06,PCPS07,LK07,PA08,SMX09,MP10}.

Entanglement has been studied for various exotic forms of entangled coherent states.
Both the Greenberger-Horne-Zeilinger type of entangled coherent states~\cite{Wan01,An03,An04,JA06,PCPS07}
and the W type of entangled coherent states~\cite{An04,JA06,GK07,GK08} have been studied
as well as the cluster type of entangled coherent states~\cite{AK09}.

The effect of dissipation and decoherence on entanglement and nonlocality is also a subject of intensive investigation for all types of 
entangled coherent states, including the robustness or fragility of the entanglement~\cite{WJK02}.
Characterization of probabilistic teleportation of coherent states via entangled coherent state quantum channel in an open system~\cite{LX03}.
Non-Markovian decoherence dynamics is important for entangled coherent states, and An, Feng and Zhang
obtain an exact master equation with and without environmental memory using influence functional theory~\cite{AFZ09}.

Strategies to mitigate decoherence of entangled coherent states, for example by squeezing~\cite{LJ09}, are of practical value.
Entanglement purification for mixed entangled coherent states is also a promising approach~\cite{JK01,CKW02}.
Park and Jeong compare the dynamics of entangled coherent states against entangled photon pair states under decoherence and inefficient detection~\cite{PJ10}. 
They discover that entangled coherent states are more robust as quantum channels for teleportation whereas entangled photon pair states are better with respect to photodetection inefficiency.

\section{Applications and Implications}
\label{sec:QIP}

Entangled coherent states have several applications as discussed earlier in this review.
For example entangled coherent states can serve as a resource for quantum teleportation~\cite{Wan01,vEH01,JKL01}
or for quantum networks~\cite{vLLMN08,EHN11}.
A `cat state' superposition of two coherent states readily serves as a qubit for quantum logical encoding~\cite{CMM99}.

The `cat state' qubit also serves as the logical basis for performing universal quantum computation~\cite{dOM00},
and entangled coherent states play an important role in such quantum information processing~\cite{JK02,JK02}.
In particular this encoding leads to entangled qubits (ebits) corresponding to entangled coherent states~\cite{MMS00}.

Entangled coherent states also serve an important role in quantum metrology,
which harnesses quantum resources such as entanglement to surpass the standard quantum limit
(due to partition noise in particle interferometry, which applies to atomic clocks and displacement measurements inter alia)~\cite{YMK86}.
Multimode even/odd coherent states are especially amenable for quantum metrology~\cite{ADM+94}.
Entangled coherent states are known to outperform other popular two-mode entangled states in quantum metrology~\cite{JMS11,JMS11E},
but perhaps its benefit is strongest for digital parameter discrimination~\cite{HKM11}.

The entangled coherent state representation~\cite{SBRK03} plays a key role in resolving fundamental issues concerning superselection 
of angular momentum~\cite{AS67}, charge~\cite{AS67} and phase~\cite{Mol97,Gea98,Mol98,RS01,vEF02,SBRK03,Wis04,Smo04}.
Essentially the entangled coherent state representation captures, 
in a mathematically simple and conceptually appealing way,
how superselection can be obviated by adding an extra degree of freedom and splitting the state to provide a reference frame.

\section{Summary and Conclusions}
\label{sec:summary}

This review provides a comprehensive summary of results concerning entangled coherent states and their generalizations
since the inception of entangled coherent states by Aharonov and Susskind in 1967 to obviate superselection.
Coherent states are appealing for their mathematical elegance as representations and closeness to classical physical states,
and entangled coherent states  build on these elegant representation properties.

Furthermore entangled coherent states have a richness due to
entanglement between these seemingly classical coherent states.
Entangled coherent states have many beautiful nonclassical properties and generalize beyond the Heisenberg-Weyl algebra of harmonic 
oscillators to the cases of spin, squeezing, pair coherent states and beyond.

Remarkably entangled coherent states have been created and observed experimentally.
These exquisitely fragile states can be manifested in the laboratory given sufficient guile.
Until now the one successful experimental realization relies on parametric amplification in two modes and photon subtraction.
Other realizations could be possible if large low-loss Kerr nonlinearities are created for propagating or for cavity fields.
Ion traps could also be promising for realizing entangled coherent states between vibrational modes,
and nanotechnology could open new vistas for entangling coherent states of motion.

Multipartite entanglement is a vast topic of research,
and entangled coherent states play an important role in this area.
Various multipartite entangled states such as Greenberger-Horne-Zeilinger, W and cluster states are studied for their rich properties 
and applications, and each of these states has nontrivial analogues with entangled coherent states.

In summary entangled coherent states have been important from superselection arguments in 1967 to 
today's applications to quantum information processing.
This review article can serve as a resource to propel studies and applications of entangled coherent states for upcoming decades,
which hold further revelations and surprises.

\acknowledgments

This project has been supported by AITF, CIFAR, NSERC, MITACS and PIMS.

\bibliography{ecs}

\end{document}